\documentclass{emulateapj}

\begin{document}

\title{A Pilot Search for Population III Supernova Candidates in the \emph{Spitzer}/IRAC Dark Field}
\author{Mark I. Frost,$^{1,3}$ Jason Surace,$^1$ Leonidas A. Moustakas,$^2$ Jessica Krick,$ ^1$}
\affil{$^1$Spitzer Science Center,  MS 220-6, Caltech, Pasadena, CA, 91125\\
$^2$JPL/Caltech, 4800 Oak Grove Dr, MS 169-327, Pasadena, CA  91109\\
$^3$Astronomy Center, University of Sussex, Brighton, England, BN1 9QH, UK\\
}

\begin{abstract}
We have undertaken a systematic search for candidate supernovae from high-redshift Population III stars in a field that has been observed with repeated imaging on a cadence of 2--3 weeks over a 2.2 year baseline, the  \emph{Spitzer}/IRAC Dark Field.  The individual epochs reach a typical 5\,$\sigma$ depth of 1\,$\mu$Jy in IRAC Channel 1 (3.6\,$\mu$m). Requiring a minimum of four epochs coverage, the total effective area searched is 214\,arcminutes$^2$.  The unprecedented depth and multi-epochal nature of these data make it ideal for a first foray to detect transient objects which may be candidate luminous Pair Instability Supernovae from the primordial-metallicity first stars.  The search was conducted over a broad range of timescales, allowing for different durations of the putative candidates' light curve plateau phases.  All candidates were vetted by inspection of the \emph{Spitzer} imaging data, as well as deep \emph{HST}/ACS F814W imaging available over the full field.  While many resolved-source objects were identified with \emph{Spitzer} variability, no transient objects were identified that could plausibly be identified as high-redshift supernovae candidates.  The resulting 95\% confidence level upper limit is 23\,deg$^{-2}$\,yr$^{-1}$,  for sources with plateau timescales under 400/(1+z) days and brightnesses above $\sim1\,\mu$Jy.
\end{abstract}


\section{Introduction}
Primordial metallicity Population III stars are thought to be the first luminous objects to form in the Universe.  Their formation marks the end of the cosmic dark ages and an important transition of the universe from a homogeneous state to a highly structured one.  The UV photons produced by such stars at high redshifts are also thought to be at least partly responsible for re-ionizing the universe (\citealt{tumlinson00}, \citealt{bromm01b}, \citealt{schaerer02}, \citealt{schaerer03}).  It is believed that the explosive events that mark the end state of such stars seed the intergalactic medium with heavy elements (\citealt{gnedin97}, \citealt{furlanetto03}, \citealt{greif07}).  Hence, studying these objects is of great importance in helping us to understand the high-redshift universe.  

To date, the supernovae marking the deaths of first stars (Pop III SNe) at high redshift have not been observed (though see \citealt{woo07}).  We turn to theoretical modeling to gain a better understanding of the properties of such stars and when they might have existed.   It is thought that Population III stars formed out of primordial-abundance H/He gas in low-mass dark matter halos.  For primordial-abundance stars it is expected that the explosion mechanism may drive not only ``classical'' supernovae, but also ``hypernovae'' for certain progenitor masses, driven through an electron-positron instability mechanism that results in explosive events with up to approximately one hundred times greater luminosities \citep{um02}.  Exactly when the epoch of the first stars began is still a matter of debate but estimates place it at $10<z<50$ (\citealt{wiseabel05}).  Thereafter, such objects could exist in primordial-metallicity pockets even at relatively low redshifts, even $z\leq2.5$ (e.g. \citealt{scann05}, \citealt{torn07}).  The primordial metallicity of these stars leads to inefficient cooling mechanisms through H$_{2}$, leading to very high stellar masses and to a top-heavy initial mass function (IMF) with a large fraction of stars having $M_{\star}>100 M_{\sun}$ (\citealt{bromm99}, \citealt{bromm02}, \citealt{abel00}, \citealt{abel02}) and possibly even greater masses (\citealt{brommloeb04}, \citealt{omukai03}).

In this paper we describe a search for candidate supernovae from high-redshift Population III stars in the \emph{Spitzer} IRAC dark field.  The target field is described in Section \ref{sec:df}.  In Section \ref{sec:expec} we summarize the theoretical modeling used as a guide for detecting transient objects, which would include candidate Pop III SNe. Section \ref{sec:method} outlines the search methods used.  In Section \ref{sec:dis} we compare the Pop III SNe rate upper limits we find with expectations from the literature, and highlight how this type of search may be extended in the future.  Where needed, we adopt $H_0$=73 km\,s$^{-1}$\,Mpc$^{-1}$, $\Omega_m=0.27$, and $\Omega_\Lambda=0.73$.

\section{The IRAC Dark Field}
\label{sec:df}
The IRAC dark field is the dark current calibration target for the Infra-Red Array Camera (IRAC) (\citealt{fazio04}) on board the \emph{Spitzer} space telescope.  It is an extragalactic field of very low background, in \emph{Spitzer}'s continuous viewing zone near the North Ecliptic Pole.  This area is observed at the start and end of each IRAC observing campaign (2--3 weeks apart) since \emph{Spitzer}'s first light.  For technical reasons anchored on the need for high-quality dark frames and on the normal precession of the observatory, there is only a modest overlap in the observed area on the sky from epoch to epoch, making for a position-dependent observed cadence in the time domain data.  

The data used in this analysis are based on 128 distinct epochs over the first 2.2 years of \emph{Spitzer}'s operations.  Each epoch is composed of multiple individual exposures at all of the available IRAC exposure times.  The total full IRAC mosaic in the adopted dataset has a total observing time of $>500$ hours, and covers an approximately circular area, 20\,arcminutes in diameter. Each point in this mosaic typically has more than $10$ hours of total integration time, with a maximum of $\sim80$ hours in the area of maximal overlap across epochs.  It is classically confusion-limited in all four IRAC channels ($3.6\mu m$, $4.5\mu m$, $5.8\mu m$, $8.0\mu m$).  The IRAC source extraction was done using Source EXtractor, \cite{bertin96}. 

The unprecedented depth and multi-epochal nature of these data make it ideal for a first foray into trying to detect supernovae from the first stars.  The field will, in fact, continue to be observed through the full extent of the \emph{Spitzer} Warm Mission, eventually giving a baseline of around seven (and possibly ten) years.  There is now a wealth of multiwavelength data available in the dark field including \emph{Chandra} and \emph{HST}/ACS F814W imaging \citep{krick08}.

To find transient objects, a common method is to search for significant residuals when differencing registered time-series observations.  However, because the IRAC point spread function is asymmetric and fairly complex, and indeed has a different orientation on the image of each epoch, it is difficult to distinguish candidate transient objects from artifacts in the difference images.  Therefore, the search is conducted through cross-correlating catalogs of objects detected in each individual epoch.  Each epoch of observation has a typical 5\,$\sigma$ depth of 1\,$\mu$Jy, or $m_{\rm 3.6\mu m}(\rm{AB}) \sim 23.9$.  The catalogs from all epochs were cross-matched with each other using a $1$\,arcsec radius, resulting in a master database with 31,492 sources, each of which had at least one detection in a single distinct epoch.  This master database contains all the critical information for a transience search.  For each object, the following are recorded: in which epochs the object is within the observed field and detected, and its flux density; in which epochs the object is within the observed field and \emph{not} detected, and the known sensitivity at that position (based on the integration time).  

If the spectral energy distributions of putative high-redshift Pop III SNe are above the measured detection limits for some period of time contained within the timespan of this survey, they will appear as transient objects.  We explore the expectations from theory and the potential efficacy of our search in the next section.  

\section{Expectations from Theory}
\label{sec:expec}
Massive Pop III stars with primordial metallicity are thought to be common at high redshift.  Stars with $M_{\star}<140 M_{\odot}$ or $M_{\star}>260 M_{\odot}$ are thought to form black holes at the end of their evolution (e.g. \citealt{fryer01}, \citealt{hegerwoosley02}).  Those which have masses between 140--260 $M_{\odot}$ are thought to end their lives as pair-instability supernova (PISNe).  Once helium burning in the core of such stars has ceased there is sufficient entropy to create positron-electron pairs \citep{wiseabel05}. This process converts thermal energy to the mass of the particle pair and the pressure in the core is reduced. This leads to a partial collapse which triggers a thermonuclear explosion. The star is completely destroyed leading to a PISN, in which no remnant is left behind.  At least one relatively local analog may already have been observed (SN~2006gy in NGC~1260; \citealt{smith07}, \citealt{woo07}), which lends support to the possibility of this mechanism.  PISNe would be ``host-less'' and as much as one hundred times more luminous than more-typical supernovae. 

To estimate the chances of detecting such objects, we turn to predictions from the literature for anticipated luminosities, durations (through their light curves), and frequency of events.  \citet{scann05}, using light curves calculated by \citet{weinmann04}, predict peak apparent magnitudes of $m_{\rm AB}\sim26.8$ at $z=10$ for 250\,$M_{\odot}$ PISN, assuming negligible extinction.  This suggests that the typical by-epoch depth of $m_{\rm 3.6\mu m}(\rm{AB}) \sim 24$ of our search may be able to detect such objects at $z\sim3-5$.  

PISN light curves are calculated in \citet{wiseabel05}, \citet{heger02}, and \citet{scann05}.  A broad plateau phase is expected, which could last from $\sim$10 days to as long as a full year in the frame of the event.  Since in the observed frame the light curve is stretched by a factor of (1$+$z), there could be events that would appear as near-continuous sources over the entire 2.2-year monitoring span of the current dataset.  As we ultimately anticipate a 7- to 10-year dataset by the end of the \emph{Spitzer} Warm Mission, there is great potential in future analyses to encompass more plateau-duration possibilities.

In Table \ref{tab:rates} we list several predicted Pop III SNe differential rates from the literature.  The rates quoted in Table \ref{tab:rates} for \citet{heger02} and \citet{cen03} incorporate the corrections determined by \cite{weinmann04} of $(1+z)^{-2}$ and $(1+z)^{-1}$, respectively. These rates are over redshift ranges largely beyond the sensitivity of our search, but are quoted here for completeness.  By the review in \citet{weinmann04}, realistic rates are expected to be $dN/dz\sim4$\,deg$^{-2}$yr$^{-1}$ for $z>15$ and 0.2\,deg$^{-2}$yr$^{-1}$ for $z>25$.  \cite{wiseabel05} find a Pop III SNe rate of 0.34\,deg$^{-2}$yr$^{-1}$ at $z=10$, which changes negligibly over the mass range $100 M_{\odot}<M_{\star}<500 M_{\odot}$.  A wide range of values are expected, $\sim$0.2$-$50\,deg$^{-2}$ yr$^{-1}$, which is indicative of how the parameters involved in such predictions are still not well-constrained.  We also refer the reader to \citet{scann05} for further discussion of these predictions. 

\begin{table}
\begin{center}
 \caption{Predicted Population III supernova rates found in the literature.}
\label{tab:rates}
\begin{tabular}{cccc} \hline
dN/dz & M$_{\rm progenitor}$ & z & Reference  \\ 
(${\rm deg}^{-2} {\rm yr}^{-1}$)& ($M_{\odot}$)  &  &    \\ \hline
$\sim$0.2&250&$\sim10$&\cite{heger02}\footnote{Rate of \cite{heger02} incorporating correction factor determined by \cite{weinmann04} and recalculated at $z=10$}\\
 0.34&100-500&10&\cite{wiseabel05}\\
50&250&$>15$&\cite{mackey03}\\
11&100&$>$13&\cite{cen03}\footnote{Corrected rate determined by \cite{weinmann04}}\\
25&140-260&5&\cite{weinmann04}\\
  \hline
\end{tabular}
\end{center}
\end{table}

\vspace{2cm}

\section{Search Method}
\label{sec:method}
The search for transient objects is based on the master database described in Section \ref{sec:df}.  Systematic searches were conducted for three different ranges of possible plateau durations (in the observed frame), which are appropriate to the 2.2-year baseline of the present dataset, and which are plausible based on some of the theoretical expectations for the light curves. These are  $0<t_{obs}<100$, $100<t_{obs}<200$, and $200<t_{obs}<400$ days.  In addition to the practical convenience of analyzing these data in this comparamentalized fashion, it also anticipates the possibility that some transient-source light curves may be truncated by either end of our 2.2 year baseline by different timescales.  

We remind the reader that the light curves of all sources are largely discontiguous because the precise pointing (and orientation) varies epoch by epoch. This also affects the precise depth of each epoch, and changes the on-sky orientation of the IRAC diffraction pattern essentially leading to rotating diffraction spikes across images.  We find that the photometry of objects near bright stars is systematically contaminated by these rotating diffraction spikes.  Therefore we identify the stars and the sources close enough to them to be affected, and explicitly remove them from our master catalog. Imaging with \emph{HST}/ACS F814W was obtained between November \& December 2006, one year after the end of the IRAC dataset described here.  With this data, a plot of the Source EXtractor \citep{bertin96} {\sc isoarea} value versus aperture magnitude photometry clearly separates the ridge of unresolved sources from resolved galaxies, because unresolved sources have smaller {\sc isoarea} than galaxies at any given magnitude.  This method identifies 1147 unresolved sources.  We empirically determine a radius of $r=0.06\times S_{3.6\mu m}$\,arcsec (but no greater than 30\,arcsec), to mask out a circular area around each star also removing an additional 5858 sources which were in close proximity. 

Given the relative non-homogeneity in the survey's depth and cadence sampling, we apply several criteria to remove erroneous sources.  1) We impose an \emph{upper} flux density limit of 40\,$\mu$Jy, which is an empirically determined practical threshold for removing additional objects that are affected by diffraction spike artifacts, beyond the masked radii, and object de-blending issues.  2)  For all sources we only include epochs which have a 5\,$\sigma$ detection of $>$1\,$\mu$Jy.  3) We do not consider sources that appear in only a \emph{single} epoch.  4) Sources detected in only two or three epochs are only included in our search if they have 5\,$\sigma$ detections of $>$1\,$\mu$Jy in \emph{all} epochs.  5) For any object with multiple epoch detections, we require the ${3.6\mu m}$ photometric uncertainty to be less than 10\% in at least half of its detected epochs.  This ensures that this source is not rejected on the basis of large uncertainties in a small number of epochs (e.g.~due to some epochs being particularly shallow relative to the others).  6) Finally, we require at least two significant non-detections, to ensure a clear transient signal. 

After application of each of these criteria, 650 candidates remained that warranted more careful follow-up.  The ${3.6\mu m}$ light curves and the corresponding IRAC and \emph{HST} imaging were all visually inspected by custom-built software that clearly identified all salient aspects for the object in the master database.  No a priori restrictions were placed on the shape of the light curve.  Several objects with otherwise flat light curves were found to have a dramatic ``flare-up''  in a single epoch.  Careful inspection of the individual IRAC 100\,s exposures of that epoch showed the spike to be a cosmic ray incident.  The remaining candidates were found to have resolved-source counterparts in the \emph{HST} data, and are therefore candidate low-redshift active galactic nuclei, which though extremely interesting in their own right, are clearly not Pop III SNe candidates.   At the end of this careful analysis and vetting procedure, \emph{no} viable candidates survived.  In the next Section we calculate the formal limits implied by our search. 

\section{Discussion \& Conclusions}
\label{sec:dis}
No Pop III SNe candidates to our sensitivity limit of $m_{\rm 3.6\mu m}(\rm AB) \sim 24$ were identified.  For the areal search over a total of 214\,arcminutes$^2$, the rate limit is below 8\,deg$^{-2}$\,yr$^{-1}$.  Using the same area but also modeling this non-detection limit as a Poisson distribution, the 95\,\%-confidence upper limit is 23\,deg$^{-2}$\,yr$^{-1}$, ignoring cosmic variance uncertainties.  At $z\leq$[3, 5, 10] this also corresponds to a volumetric rate of  [910, 480, 270]\,Gpc$^{-3}$\,yr$^{-1}$, respectively.  These limits are approximate, as the precise survey area relevant to each individual source is a complicated function of the varying field and depth by epoch.  Furthermore, these limits only apply to moderate-duration events, with plateau phases lasting less than $\sim400/(1+z)$\,days, by construction of the search.  Many of the predicted light curves given e.g.\ by \citet{scann05} have plateau phases up to 1\,yr in the rest frame, which would be too long in the observed frame to be detected by the present survey, but which may be approachable in future incarnations of the \emph{Spitzer} Dark Field surveys.  

The \citet{wiseabel05} and \citet{heger02} $z=10$ predicted differential rates of $dN/dz\sim0.34$\,deg$^{-2}$\,yr$^{-1}$ and $\sim$0.2\,deg$^{-2}$\,yr$^{-1}$, respectively, are not ruled out, even if they appear at lower redshifts which may be above our sensitivity limit.  Given that our search is most effectively probing $z\sim3-5$, the differential rate of \cite{weinmann04} of 25 \,deg$^{-2}$\,yr$^{-1}$ at $z=5$ is broadly comparable to our limit.  It should be noted the very luminous PISNe that we would be sensitive to in this survey may be only a small fraction of all high-redshift supernova events.  There is a distinction currently being made between Pop III.1 and III.2 stars, where the former class are of fully primordial abundance, and form in dark matter mini-halos, resulting in the pre-requisite stellar masses of above $\sim100\,M_{\odot}$\, to produce PISNe (\citealt{jobro06}, \citealt{mcktan08}). These are distinct from the Pop III.2 stars, which are expected to form through atomic cooling processes, producing only $\sim 10\,M_{\odot}$ progenitors \citep{grebro06}.  The PISNe Pop III.1 progenitors may consist of only some 10\% of Pop III SNe \citep{grebro06}.  Furthermore, the ``pristine'' Pop III.1 progenitors can suffer dramatic negative feedback (e.g.\ \citealt{mcktan08}), which may additionally limit their relative numbers.  These were all considerations not yet taken in the predictions by \citet{mackey03}, which led to the high expectation rates.  

To do a \emph{comprehensive} search for Pop III SNe using \emph{Spitzer} (or a future platform with similar capabilities) would require a survey area of 1 deg$^2$ with exposures of 5000 seconds ($50\times100$\,s exposures) per point on the sky. This would reach $m_{\rm AB}\sim26$ at the 3-5\,$\sigma$ level, depending on extinction.  Each epoch would require approximately 200 hours of observation to cover the survey area with a 5 arcminute field of view. This greater depth complicates the IRAC data reduction as confusion is a significant, but tractable problem. Trading depth for greater area may not be optimal for detecting the \emph{highest}-redshift  Pop III SNe, which are expected to be quite faint, $m_{\rm AB}\gtrsim26$.  This requirement is increasingly relaxed for lower redshift events, which may possibly exist at redshifts as low as $z\sim2$.  This ideal survey would need to be imaged every two months for a period of several years in order to plausibly sample much or all of the plateau phase of the light curve. The total program would therefore be of the order of 2000-3000 hours. 

In the meanwhile, the current dataset continues to expand as \emph{Spitzer} continues to operate, acquiring new IRAC observations every 2--3 weeks.  Including observations made during the \emph{Spitzer} Warm Mission, the full dataset will span at least seven years, and maybe as many as ten full years.  Future searches following the procedure described here should be powerful for identifying plausible candidates, or at least in setting firmer limits on the production rate of PISNe at high redshift, and thus setting practical limits on the relative abundances of Pop III.1 versus Pop III.2 progenitors ultimately informing studies of reionization of the high-redshift universe. 


\acknowledgements
Support for this work was provided by NASA through the \emph{Spitzer} Space Telescope Visiting Graduate Student Program, through a contract issued by the Jet Propulsion Laboratory, California Institute of Technology under a contract with NASA.  The work of LAM was carried out at the Jet Propulsion Laboratory, California Institute of Technology, under a contract with NASA. This research has made use of data from the {\it Spitzer} Space Telescope, which is operated by the Jet Propulsion Laboratory, California Institute of Technology under a contract with NASA, and the NASA/ESA {\it Hubble} Space Telescope, obtained at the Space Telescope Science Institute, which is operated by the Association of Universities for Research in Astronomy, Inc., under NASA contract NAS 5-26555. These observations are associated with program \#10521. Support for program \#10521 was provided by NASA through a grant from the Space Telescope Science Institute, which is operated by the Association of Universities for Research in Astronomy,Inc., under NASA contract NAS 5-26555. Support for this work was also provided by STFC studentship PPA/S/S2005/04270.   LAM is grateful to Tom Abel and Ranga-Ram Chary for many discussions on this topic. We would also like to thank Seb Oliver for his useful comments and suggestions.  We are grateful to the anonymous referee for comments that have helped focus and improve this report. 

\bibliographystyle{apj}

\begin{thebibliography}{}

\bibitem[Abel et al.(2000)]{abel00} Abel, T., Bryan, G.~L., 
\& Norman, M.~L.\ 2000, \apj, 540, 39 

\bibitem[Abel et al.(2002)]{abel02} Abel, T., Bryan, G.~L., 
\& Norman, M.~L.\ 2002, Science, 295, 93

\bibitem[Bertin 
\& Arnouts(1996)]{bertin96} Bertin, E., \& Arnouts, S.\ 1996, \aaps, 117, 393

\bibitem[Bromm et al.(1999)]{bromm99} Bromm, V., Coppi, P.~S., 
\& Larson, R.~B.\ 1999, \apjl, 527, L5 

\bibitem[Bromm et al.(2001)]{bromm01b} Bromm, V., Kudritzki, 
R.~P., \& Loeb, A.\ 2001, \apj, 552, 464 

\bibitem[Bromm et al.(2002)]{bromm02} Bromm, V., Coppi, P.~S., \& Larson, R.~B.\ 2002, \apj, 564, 23 

\bibitem[Bromm 
\& Loeb(2004)]{brommloeb04} Bromm, V., \& Loeb, A.\ 2004, New Astronomy, 9, 353 

\bibitem[Cen(2003)]{cen03} Cen, R.\ 2003, \apj, 591, 12 

\bibitem[Fazio et al.(2004)]{fazio04} Fazio, G.~G., et al.\ 
2004, \apjs, 154, 10 

\bibitem[Fryer et al.(2001)]{fryer01} Fryer, C.~L., Woosley, 
S.~E., \& Heger, A.\ 2001, \apj, 550, 372 

\bibitem[Furlanetto 
\& Loeb(2003)]{furlanetto03} Furlanetto, S.~R., \& Loeb, A.\ 2003, \apj, 588, 18 

\bibitem[Gnedin 
\& Ostriker(1997)]{gnedin97} Gnedin, N.~Y., \& Ostriker, J.~P.\ 1997, \apj, 486, 581 

\bibitem[Greif \& Bromm(2006)]{grebro06} Greif, T., \& Bromm, V.\ 2006, MNRAS, 373, 128

\bibitem[Greif et al.(2007)]{greif07} Greif, T., Johnson, J.~L., Bromm, V., Klessen, R.~S.\ 2007, \apj, 670, 1

\bibitem[Heger 
\& Woosley(2002)]{hegerwoosley02} Heger, A., \& Woosley, S.~E.\ 2002, \apj, 567, 532 

\bibitem[Heger et al.(2002)]{heger02} Heger, A., Woosley, S., 
Baraffe, I., 
\& Abel, T.\ 2002, Lighthouses of the Universe: The Most Luminous Celestial Objects and Their Use for Cosmology: Proceedings of the MPA/ESO/MPE/USM Joint Astronomy Conference Held in Garching, Germany, 6-10 August 2001, ESO ASTROPHYSICS SYMPOSIA.~ISBN 3-540-43769-X.~Edited by M.~Gilfanov, R.~Sunyaev, and E.~Churazov.~Springer-Verlag, 2002, p.~369, 369 

\bibitem[Johnson \& Bromm(2006)]{jobro06} Johnson, J.~L., \& Bromm, V.\ 2006, MNRAS, 366, 247

\bibitem[Krick et al.(2008)]{krick08} Krick, J.~E., Surace, J.~A., Thompson, D., Ashby, M.~L.~N., Hora, J.~L., Gorjian, V., \& Yan, L.\ 2008, \apj, 686, 918

\bibitem[Mackey et al.(2003)]{mackey03} Mackey, J., Bromm, V., 
\& Hernquist, L.\ 2003, \apj, 586, 1 

\bibitem[McKee \& Tan(2008)]{mcktan08} McKey, C.~F., \& Tan, J.~C.\ 2008, ApJ, 681, 771

\bibitem[Miralda-Escude 
\& Rees(1997)]{miralda97} Miralda-Escude, J., \& Rees, M.~J.\ 1997, \apjl, 478, L57 

\bibitem[Omukai 
\& Palla(2003)]{omukai03} Omukai, K., \& Palla, F.\ 2003, \apj, 589, 677 

\bibitem[O'Shea et al.(2008)]{oshea08} O'Shea, B.~W., McKee, 
C.~F., Heger, A., \& Abel, T.\ 2008, ArXiv e-prints, 801, arXiv:0801.2124 

\bibitem[Scannapieco(2005)]{scann05} Scannapieco, E., Madau, P., Woosley, S., Heger, A., \& Ferrara, A.\ 2005, \apj, 633, 1031

\bibitem[Schaerer(2002)]{schaerer02} Schaerer, D.\ 2002, \aap, 382, 28

\bibitem[Schaerer(2003)]{schaerer03} Schaerer, D.\ 2003, \aap, 397, 527 

\bibitem[Smith et al.(2007)]{smith07} Smith, N., et al.\ 2007, \apj, 666, 1116

\bibitem[Tornatore et al.(2007)]{torn07} Tornatore, L., Ferrara, A., \& Schneider, R.\ 2007, MNRAS, 382, 945 

\bibitem[Tumlinson 
\& Shull(2000)]{tumlinson00} Tumlinson, J., \& Shull, J.~M.\ 2000, \apjl, 528, L65 

\bibitem[Umeda \& Nomoto(2002)]{um02} Umeda, H., \& Nomoto, K.\ 2002, \apj, 565, 385

\bibitem[Weinmann 
\& Lilly(2005)]{weinmann04} Weinmann, S.~M., \& Lilly, S.~J.\ 2005, \apj, 624, 526 

\bibitem[Wise 
\& Abel(2003)]{wiseabel03} Wise, J.~H., \& Abel, T.\ 2003, The Emergence of Cosmic Structure, 666, 97 

\bibitem[Wise 
\& Abel(2005)]{wiseabel05} Wise, J.~H., \& Abel, T.\ 2005, \apj, 629, 615 

\bibitem[Woosley et al.(2007)]{woo07} Woosley, S.~E., Blinnikov, S., \& Heger, A.\ 2007, Nature, 450, 7168

\bibitem[Yoshida et al.(2006)]{yoshida06} Yoshida, N., Omukai, 
K., Hernquist, L., \& Abel, T.\ 2006, \apj, 652, 6 
  
 \end{thebibliography}

\label{lastpage}

\end{document}